\documentclass[11pt]{article}
\usepackage{graphicx}
\usepackage{epsfig}
\usepackage{array}
\usepackage[figuresright]{rotating}
\usepackage{amsmath}
\usepackage{amssymb}
\usepackage{amsfonts}

\newcommand{\pibf}{\mbox{\boldmath $\pi$}}
\newcommand{\taubf}{\mbox{\boldmath $\tau$}}

 \newcommand{\sigbf}{\mbox{\boldmath $\sigma$}}

\newcommand{\veceps}{\mbox{\boldmath$\epsilon$}}

\newcommand{\vectau}{\mbox{\boldmath$\tau$}}

\newcommand{\fdagger}{\mbox{$/\!\!\!\partial$}}
\usepackage{cite}
\begin{document}
\renewcommand{\thefootnote}{\fnsymbol{footnote}}
\begin{center}
{\LARGE{\bf Observation of the Higgs Boson of strong interaction via
Compton scattering by the nucleon\footnote{R\'esum\'e of an invited 
talk presented 
for the A2 collaboration at MAMI (Mainz)}} 
  }\\[1ex] 
Martin Schumacher\\mschuma3@gwdg.de\\
Zweites Physikalisches Institut der Universit\"at G\"ottingen,
Friedrich-Hund-Platz 1\\ D-37077 G\"ottingen, Germany\\
\end{center}
\renewcommand{\thefootnote}{\arabic{footnote}}
\begin{abstract}
It is shown that the Quark-Level Linear $\sigma$ Model (QLL$\sigma$M)
leads to a prediction for the
diamagnetic term of the polarizabilities  of the
nucleon which is in excellent agreement with experimental data. The 
bare mass of
the $\sigma$ meson  is predicted to be  $m_\sigma=666$
MeV and the two-photon width
$\Gamma(\sigma\to\gamma\gamma)= (2.6\pm 0.3)$ keV.
It is argued  that the mass predicted by the  QLL$\sigma$M corresponds to
the  $\gamma\gamma\to\sigma\to N\bar{N}$ reaction, i.e. to a $t$-channel
pole of the  $\gamma N\to N \gamma$ reaction. 
Large-angle Compton scattering experiments revealing
effects of the $\sigma$ meson in the differential cross section are discussed.
Arguments are presented that these findings may be understood as an 
observation of the Higgs boson of strong interaction while being a part 
of the constituent quark.
\end{abstract}

\section{Introduction}

The $\sigma$ meson  introduced by Schwinger  \cite{schwinger57}
and Gell-Mann--Levy \cite{gellmann60} has attracted great interest
because  there are good reasons to consider it as the Higgs boson of strong
interaction,
an aspect which recently has been
emphasized by several authors \cite{close02,tornqvist02,tornqvist05,
pennington06a}.
 The mass 
of the $\sigma$ meson  is predicted by the Quark-Level Linear $\sigma$ 
Model (QLL$\sigma$M) \cite{delbourgo95} (see also
\cite{schumacher06,beveren09}
and references therein)
to be $m_\sigma = 666$ MeV, whereas $\pi$-$\pi$  scattering analyses led to
$\sqrt{s_\sigma}=M_\sigma -i\,\Gamma_\sigma/2$ with 
$M_\sigma=441^{+16}_{-8}$ MeV and $\Gamma_\sigma=544^{+18}_{-25}$ MeV
in  one recent evaluation (CCL)\cite{caprini06}, or 
$\sqrt{s_\sigma}=(476-628)-i(226-346)$ MeV when tests of the stability of fits
to data are taken into account \cite{beveren06}.
It certainly is extremely important to understand how these two masses are
related to each other.  The finding is that 
 $m_\sigma=666$ MeV is the bare mass of the $\sigma$ meson which is  
observed  in  space-like Compton scattering 
$\gamma\gamma\to\sigma\to N\bar{N}$, i.e. as a $t$-channel pole
of Compton scattering  $\gamma N\to N \gamma$. 
For the electromagnetic polarizabilities space-like Compton scattering
is equally important as 
time-like Compton scattering  $\gamma N\to N \gamma $      ($s$-channel).
The $\sigma$ and non-$\sigma$  components of the electromagnetic 
polarizabilities for the proton $(p)$ and the neutron $(n)$
are $\alpha_{p,n}(\sigma)=-\beta_{p,n}=7.6$, 
$\alpha_p({\rm non-}\sigma)=4.4$, $\beta_p({\rm non-}\sigma)=9.5$, 
$\alpha_n({\rm non-}\sigma)=5.8$, $\beta_n({\rm non-}\sigma)=9.4$ in units of
$10^{-4}$ fm$^3$. 
On the  quark level the reaction $\gamma \gamma\to \sigma \to N\bar{N}$
implies that two photons with parallel planes of linear polarization
interact with   $\sigma$ mesons, being  parts of the constituent quarks.
In this sense it is justified to consider space-like Compton scattering 
as an in-situ observation of the Higgs boson 
of strong interaction. 

In addition to the specific  aspects of the $\sigma$ meson as outlined in 
the preceding paragraph this particle  is of interest because it 
has been observed as an intermediate
state in many reactions \cite{PDG}.
Therefore,  there is no doubt that this particle exists and that it
belongs to a scalar nonet ($\sigma(600)$, $f_0(980)$,   $a_0(980)$,
 $\kappa(800)$) below 1 GeV. A problem may appear due to the fact that there
is also a scalar nonet above 1 GeV.  This has led to the assumption
that the scalar nonet above 1 GeV should be understood as
$q\bar{q}$ states whereas  the scalar nonet below 1 GeV should be understood
as $qq\bar{q}\bar{q}$ states (see e.g. \cite{close02} and references
therein). Other versions  consider 
meson molecules and gluonic components 
(see \cite{tornqvist02,close02,tornqvist05,pennington06a,pennington06b,pennington07a,klempt07,pennington07b} 
and references therein). 
It is hard to see  that strict
criteria can be found giving proof of the validity of one model and
excluding the validity of an other. The best way to proceed is to start with
a model for the low-mass scalar mesons where $q\bar{q}$  is a $^3P_0$ 
core state which may couple to other hadronic or gluonic 
configurations \cite{klempt07}. Then the essential properties as e.g. 
the two-photon
width $\Gamma(\sigma\to\gamma\gamma)$ may be understood in terms of the 
$q\bar{q}$
core   whereas  other components may show up in hadronic reactions 
where scalar mesons appear  as intermediate states.

The present work is a continuation of a systematic series of studies 
\cite{schumacher06,schumacher07a,schumacher07b,schumacher08,schumacher09}
on the 
electromagnetic structure of the nucleon, following experimental work
on Compton scattering and  a comprehensive  review on this topic
\cite{schumacher05}. These recent  investigations  have shown
\cite{schumacher06,schumacher07a,schumacher07b,schumacher08,schumacher09}
that a  systematic study of all partial resonant and nonresonant
photo-excitation processes of the nucleon and of their relevance for the
fundamental structure constants of the nucleon 
as there are the electric
polarizability ($\alpha$), the magnetic polarizability ($\beta$) and the
backward spin-polarizability ($\gamma_\pi$)
is essential for
an understanding of the electromagnetic structure of the nucleon. 
In addition it has been found that
the structure of the constituent quarks and their
coupling 
to pseudoscalar and scalar mesons is  important for 
the understanding of the electric and 
magnetic polarizabilities 
and of the backward spin-polarizability. The main purpose of 
the present  work is to
prove that the method of calculating the $t$-channel contribution from the 
reaction $\gamma\gamma\to\sigma\to N\bar{N}$ where the properties of the
$\sigma$ meson are  taken from the QLL$\sigma$M is a precise procedure
and largely superior to previous approaches where the combination 
of the two reactions $\gamma\gamma\to\sigma\to\pi\pi$ and 
$\gamma\gamma\to\sigma\to N\bar{N}$ is exploited.

\section{The dynamical linear $\sigma$ model on the quark level and the mass
  of the $\sigma$-meson \label{Thedynamical}}

In the following we give a r\'esum\'e of the 
 dynamical linear $\sigma$ model on the quark level \cite{delbourgo95}
which in short term
may be named Quark-Level Linear $\sigma$ Model (QLL$\sigma$M) 
\cite{beveren09}.

The Lagrangian of the Nambu--Jona-Lasinio (NJL) model 
has been formulated in two equivalent ways 
\cite{lurie64,eguchi76,vogl91,klevansky92}
\begin{eqnarray}
&&{\cal L}_{\rm NJL}=\bar{\psi}(i\fdagger-m_0) \psi
+ \frac{G}{2}[(\bar{\psi}\psi)^2+(\bar{\psi}i\gamma_5\vectau\psi)^2],
\label{NJL1}\\
{\rm and}\hspace{1cm} && \nonumber\\
&&{\cal L'}_ {\rm
  NJL}=\bar{\psi}i\fdagger\psi-g\bar{\psi}(\sigma+i\gamma_ 5
\vectau\cdot\pibf)\psi-\frac12\delta\mu^2(\sigma^2+\pibf^2)+\frac{gm_0}{G}
\sigma,
\label{NJL2}\\
{\rm where} \hspace{1cm}&&\nonumber\\
&&G=g^2/\delta\mu^2\quad \mbox{and}\quad \delta\mu^2=(m^{\rm cl}_\sigma)^2.
\label{grelations}
\end{eqnarray}
Eq. (\ref{NJL1}) describes the four-fermion version of the
NJL model and Eq.  (\ref{NJL2}) the bosonized  version. 
The quantity $m_0=(m_u+m_d)/2$ denotes the average current quark mass for two
flavors,
$\psi$  the spinor of constituent quarks with two
flavors. The quantity $G$ is the coupling constant of the four-fermion
version, $g$ the Yukawa coupling constant and $\delta\mu$ a mass parameter
entering into the mass counter-term of Eq. (\ref{NJL2}). The coupling
constants $G$, $g$ and  the mass parameter $\delta\mu$ 
are related to each other 
and to the $\sigma$ meson mass in the chiral limit (cl),
$m^{\rm cl}_\sigma$, as
given in (\ref{grelations}).  The relation $\delta\mu^2=(m^{\rm
    cl}_\sigma)^2$ can easily be derived by applying spontaneous symmetry
  breaking to $\delta\mu$ in analogy 
to spontaneous symmetry breaking predicted by the linear $\sigma$ model
  (L$\sigma$M) for the mass parameter $\mu$ entering into that model
\cite{dealfaro73}.

Using diagrammatic techniques the following  equations  may be found 
\cite{delbourgo95,klevansky92,hatsuda94}
\begin{eqnarray}
&&M^*=m_0+ 8\, i\, G N_c \int^{\Lambda}\frac{d^4 p}{(2\pi)^4}
\frac{M^*}{p^2-M^{*2}},\quad M=-\frac{8 i N_c g^2}{(m^{\rm cl}_\sigma)^2}
\int\frac{d^4 p }{(2\pi)^4}\frac{M}{p^2-M^2},
\label{gapdiagram}\\
&&f^2_\pi = -4\,i\,N_cM^{*2} \int^{\Lambda}\frac{d^4p}{
(2\pi)^4}\frac{1}{(p^2-M^{*2})^2}, \quad
f^{\rm cl}_\pi=-4iN_cgM\int \frac{d^4p}{(2\pi)^4}\frac{1}{(p^2-M^2)^2},
\label{fpiexpress}\\
&&m^2_\pi=-\frac{m_0}{M^*}\frac{1}{4\,i\,G N_c I(m^2_\pi)},
\label{pionmass-2}\\
&&I(k^2)=\int^{\Lambda}\frac{d^4p}{(2\pi)^4}\frac{1}{[(p+\frac12 k)^2-M^{*2}][
(p-\frac12 k)^2-M^{*2}]}.
\nonumber
\end{eqnarray}
The expression given on the l.h.s. of  (\ref{gapdiagram}) 
is the gap equation with $M^*$
being the mass of the constituent quark with the contribution 
$m_0$ of the
current
quarks included and $N_c=3$ being the number of colors. 
The quantity $M$ on the r.h.s. of Eq. (\ref{gapdiagram}) is the mass
of the constituent quark in the chiral limit.
The l.h.s. of Eq. (\ref{fpiexpress}) represents
an expression for  the pion decay constant having 
the experimental value $f_\pi=(92.42\pm 0.26)$  MeV \cite{PDG}.
On the  r.h.s. of Eq. (\ref{fpiexpress}) the corresponding expression for
$f^{\rm cl}_\pi$ is given which is the   same quantity in the chiral limit.
The relation for the pion mass in  Eq. (\ref{pionmass-2})
has no  counterpart in the chiral limit because there the pion mass is
zero.

In principle the gap parameter $M^*$ and through this the mass of the 
$\sigma$ meson $m_\sigma$ can be determined using the l.h.s. relations of Eqs.
(\ref{gapdiagram})  to   (\ref{pionmass-2}). This has been carried out in
\cite{hatsuda94} leading to $m_\sigma\approx 2 M^* \approx 668$ MeV.

A more  elegant way to predict $m_\sigma$ on an absolute scale introduced by
Delbourgo and Scadron  \cite{delbourgo95}          is obtained by
exploiting the r.h.s. of Eqs. (\ref{gapdiagram}) and (\ref{fpiexpress}).
Making use of the identity (dimensional regularization 
\cite{thomas00,delbourgo95})
\begin{equation}
-i\frac{(m^{\rm cl}_\sigma)^2}{16N_cg^2}=
\int\frac{d^4 p}{(2\pi)^4}\left[\frac{M^2}{(p^2-M^2)^2}-
\frac{1}{p^2-M^2}\right]
=-\frac{iM^2}{(4\pi)^2}
\label{identity}
\end{equation}
we arrive at 
\begin{equation}
(m^{\rm cl}_\sigma)^2=\frac{N_c g^2 M^2}{\pi^2},
\label{sigmamasssquared}
\end{equation}
and with $m^{\rm cl}_\sigma=2M$ at the Delbourgo-Scadron relation 
\cite{delbourgo95}
\begin{equation}
g=g_{\pi qq}=g_{\sigma qq}= 2\pi/\sqrt{N_c}=3.63.
\label{coupling}
\end{equation}
The $\sigma$-meson mass corresponding to this coupling constant is
\begin{equation}
m_\sigma= 666.0\,\,\,{\rm MeV},
\label{sigmamassfinal-2}
\end{equation}
where use has been made of
$m^2_\sigma= (m^{\rm cl}_\sigma)^2 + {\hat m}^2_\pi$,
$f^{\rm cl}_\pi=89.8$ MeV \cite{nagy04}, $M=325.8$ MeV  and 
${\hat m}_\pi= 138.0$ MeV.  This result may also be compared with information
from magnetic moments\footnote{It should be noted that the magnetic
  moments lead to $m_\sigma \approx 2M^* \approx 664$ MeV when averaged over
  the results for the proton and the neutron (see the second reference in
\cite{schumacher08}).}.

It has to be noted that the present version of the QLL$\sigma$M is incomplete
in the sense that the coupling of the $\sigma$ meson to meson loops is not 
taken into account. These meson loops lead to  contributions to the
two-photon decay width $\Gamma(\sigma\to\gamma\gamma)$ of the $\sigma$ meson
in addition to the dominant $q\bar{q}$ contribution (see \cite{beveren09} and
Table \ref{oller5}).

\section{The two-photon decay width of the $\sigma$ meson \label{thetwophoton}}

In the quark model the $\sigma$ meson has the structure
\begin{equation}
|n\bar{n}\rangle=\frac{1}{\sqrt{2}}|u\bar{u}+d\bar{d}\rangle.
\label{nbarn}
\end{equation}
Then the principal contributions to the amplitude
${\cal M}(\sigma\to\gamma\gamma)$ comes from the up and down quark triangle
diagrams \cite{beveren09}, yielding (with $N_c=3$)
\begin{eqnarray}
&&{\cal M}(\sigma\to \gamma\gamma)=\frac{5 \alpha_e}{3\pi f_\pi}
V_q(\xi),\label{twoph}\\
&&
V_q(\xi)=2\xi[2+(1-4\xi)I(\xi)],
\label{twophotonsigma-1}
\end{eqnarray} 
where $\alpha_e=e^2/4\pi=1/137.04$, $\xi={\hat m}^2_q/m^2_\sigma$ 
and $I(\xi)$ is the triangle loop integral given by 
\begin{equation}
I(\xi)
\begin{cases}
=\frac{\pi^2}{2}-2\log^2\left[\sqrt{\frac{1}{4\xi}}+
 \sqrt{\frac{1}{4\xi}-1}\right]+ 2\pi\,i\,\log\left[\sqrt{\frac{1}{4\xi}}+
\sqrt{\frac{1}{4\xi}-1}\right]\,\,\,(\xi\leq 0.25),
\\
=2\arcsin^2\left[\sqrt{\frac{1}{4\xi}}\right]\,\,\,(\xi\geq 0.25).
\end{cases}
\label{twophotonsigma-2}
\end{equation}
With  the average constituent quark mass 
${\hat m}_q= 330$ MeV and $m_\sigma=666$ MeV we arrive at a correction
factor of
$V_q(\xi)=1.025$, whereas in the chiral limit, ${\hat m}_q\to M$ and 
$m_\sigma\to 2M$, we  have 
$V_q(\xi)=1.0$.
We see that the possible correction due to the factor $V_q(\xi)$ is small
and may be disregarded in view of possible other uncertainties. 
Then with $V_q(\xi)=1$ and 
\begin{equation}
\Gamma(\sigma\to\gamma\gamma)=\frac{m^3_\sigma}{64 \pi}|{\cal
  M}(\sigma\to\gamma\gamma)|^2
\label{gamformula}
\end{equation} 
we arrive at
\begin{equation}
\Gamma(\sigma\to\gamma\gamma)=(2.6\pm 0.3)\,\,\,\text{keV}.
\label{width}
\end{equation}
The value for the two-photon decay width given in (\ref{width}) leads to an
excellent agreement with the experimental
electromagnetic  polarizabilities  of the nucleon and, therefore, is
experimentally confirmed through this agreement. This is 
especially true for the electric polarizability $\alpha_p$ of the proton
which has the smallest experimental error, {\it{viz.}} 
$\Delta\alpha_p/\alpha_p  \approx 5 \%$. Using 
$\Delta\Gamma(\sigma\to\gamma\gamma)/\Gamma(\sigma\to\gamma\gamma)
\approx  2 \Delta\alpha_p/\alpha_p\approx 10\%$ the error given in 
(\ref{width}) is obtained (see  Table
\ref{tab} for details).

\section{Properties of the reaction $\gamma\gamma\to\sigma\to N{\bar N}$}

Let us first discuss the kinematics of Compton scattering. The conservation
of energy and momentum in nucleon Compton scattering 
\begin{equation}
\gamma(k,\lambda)+N(p)\to \gamma'(k',\lambda')+N'(p)
\label{Compton reaction}
\end{equation}
is given by
\begin{equation}
k+p=k'+p'
\label{momentumconservation}
\end{equation}
where $k$ and $k'$ are the 4-momenta of the incoming and outgoing photon,
$\lambda$ and $\lambda'$ the respective helicities and
$p$ and $p'$  the 4-momenta of the incoming and outgoing nucleon.
Mandelstam variables are introduced through the relations
\begin{eqnarray}
&&s=(k+p)^2=(k'+p')^2,\label{1}\\
&&t=(k-k')^2=(p'-p)^2,\label{2}\\
&&u=(k-p')^2=(k'-p)^2,\label{3}\\
&&s+t+u=2 m^2,\label{4}
\end{eqnarray}   
where $m$ is the mass of the nucleon. At negative $t$ this quantity can be
interpreted in terms of the c.m. scattering angle $\theta_s$:
\begin{equation}
\sin^2 \frac{\theta_s}{2}=-\frac{s t}{(s-m^2)^2}.
\label{5}
\end{equation}
Compton scattering may be described by invariant amplitudes $A_i(s,t)$
$(i=1-6)$ which
are analytic functions in the two variables $s$ and $t$  and, therefore,
may be treated in terms of dispersion relations. The third
variable $u$ is not taken into account because of the constraint given in
(\ref{4}).  The degrees of freedom of the nucleon including the structure 
of the constituent quarks enter into the invariant
amplitudes via a cut on the real axis of the complex  $s$-plane and in terms
of pointlike singularities on the positive real axis of the 
$t$-plane. The $s$-channel cut contains
the total photoabsorption cross section and through a decomposition in terms
of excitation mechanisms the complete electromagnetic 
structure of the nucleon as seen in one-photon processes. The 
pointlike singularities on the positive real $t$-axis may be related 
to the structure of the constituent quarks which couple to all mesons
with a nonzero meson-quark coupling constant. Of these the $\pi^0$ and the
$\sigma$ meson are of special interest. On the one hand they couple to two
photons with perpendicular and parallel planes of linear polarization,
respectively, and 
on the other hand they enter into the QLL$\sigma$M as described in
Section \ref{Thedynamical}.
In a formal sense the singularities on the positive real $t$-axis
correspond to the fusion of two photons with 4-momenta $k_1$ and $k_2$
and helicities $\lambda_1$ and $\lambda_2$ to form a $t$-channel intermediate
state $|\pi^0\rangle$ or $|\sigma\rangle$ from which -- in a second step --
a proton-antiproton pair is created. The corresponding reaction may be
formulated in the form
\begin{equation}
\gamma(k_1,\lambda_1)+\gamma(k_2,\lambda_2)\to \bar{N}(p_1) + N(p_2).
\label{6}
\end{equation}
In the present case the
$N\bar{N}$ pair creation-process is virtual, i.e. the energy is too low to put
the  proton-antiproton pair on the mass shell. In the c.m. system the
3-momenta of the two photons are related by ${\bf k_1}+ {\bf k_2}=0$, leading 
to $t=(\omega_1+\omega_2)^2=(W^t)^2$ where $W^t$ is the total energy.

The   $\pi^0$
singularity enters into the invariant amplitude $A_2$ and the $\sigma$ 
singularity into the invariant amplitude $A_1$ \cite{lvov97}.
 Let us first discuss
the well known case of the $\pi^0$ pole contribution. The fixed $\theta=\pi$
dispersion relation
is 
\begin{equation}
{\tilde A}_2^{\pi^0}(s,t)=\frac{1}{\pi}\int^\infty_{t_0} {\rm Im}_t 
{\tilde A}_2(s',t')
\frac{d t'}{t'-t-i0}
\label{7}
\end{equation}
with the solution for a pointlike singularity \cite{lvov99,schumacher05}
\begin{equation}
{\tilde A}^{\pi^0}_2(t)=\frac{ {\cal M}(\pi^0 \to \gamma\gamma)\, g_{\pi NN} }
{t-m^2_{\pi^0}}\taubf_3.
\label{8}
\end{equation}
The analogous formula for the $\sigma$ meson is
\begin{equation}
 {\tilde A}^{\sigma}_1(t)=\frac{ {\cal M}(\sigma \to \gamma\gamma)\, g_{\sigma NN} }
{t-m^2_{\sigma}}.
\label{9}
\end{equation}
The relation given in (\ref{8}) is  valid in the 
whole complex $t$-plane especially also in the physical range of the Compton
scattering process at $\theta=\pi$ where kinematical constraints are absent. 
For (\ref{9}) the same is true 
provided the intermediate
state in the reaction $\gamma\gamma\to\sigma\to N\bar{N}$ has 
 a definite mass of $m_\sigma= 666$ MeV as conjectured here.
The observation that the predicted mass $m_\sigma=666$ MeV together
with the predicted two-photon width $\Gamma(\sigma\to\gamma\gamma)=2.6$ keV
 obtained for the $\sigma=\frac{1}{\sqrt{2}}(u\bar{u}+d\bar{d})$ 
configuration is in perfect 
agreement with the experimental electromagnetic polarizabilities is one
of the main achievements of the present and our related previous works,
thus supporting  our supposition that this description of the  
$\sigma$ meson pole contribution  is correct.

\subsection{Analysis of the 
  $\gamma\gamma\to\sigma\to\pi\pi$ reaction}
 
Table \ref{oller5} summarizes the results available for the two-photon width 
$\Gamma(\sigma\to\gamma\gamma)$ obtained from the most recent
analyses of the  $\gamma\gamma\to\sigma\to\pi\pi$ reactions.
For comparison we also give the present result obtained
from the QLL$\sigma$M model including the $q\bar{q}$ component only
and the full calculation of \cite{beveren09} where also meson loops are
included.  From the large number of very different results recently published
for the   $\gamma\gamma\to\sigma\to\pi\pi$ reaction it is difficult
to decide whether $2.6$ or $3.5$ given in lines 3 and 4 of Table
\ref{oller5} are in better agreement with the $\gamma\gamma\to\sigma\to\pi\pi$
data. However, we should keep in mind that the result  ($2.6\pm 0.3$) keV
is supported by the experimental electric polarizability $\alpha_p$ (see Table
\ref{tab}).
\begin{table}[h]
\caption{Two-photon width of the $\sigma$ meson from the
$\gamma\gamma\to\pi\pi$ reaction compared with  the QLL$\sigma$M. In line
3  the $q\bar{q}$ contribution and information obtained from the electric
polarizability $\alpha_p$ of the proton
is taken into account. In line 4 the full 
calculation including
meson loops is taken into account.}
\begin{center}
\begin{tabular}{clll}
\hline
$\Gamma(\sigma\to\gamma\gamma)$ &method
&reference\\
$[{\rm keV}]$ & $[{\rm MeV}]$&\\
\hline
$2.6\pm 0.3$
& QLL$\sigma$M $q\bar{q}$ and $\alpha_p$& present\\
$3.5$ & QLL$\sigma$M $q\bar{q}$ and meson loops& \cite{beveren09}\\
\hline\hline
$4.1\pm 0.3$&$\gamma\gamma\to\pi^0\pi^0$
& \cite{pennington06b} (result 1)\\
$1.8\pm 0.4$&$\gamma\gamma\to\pi^0\pi^0$
& \cite{oller08} (result 2)\\
$2.1\pm 0.3$&$\gamma\gamma\to\pi^0\pi^0$
& \cite{oller08} (result 3)\\
$3.0\pm 0.3$&$\gamma\gamma\to\pi^0\pi^0$
& \cite{oller08} (result 4)\\
$1.68\pm 0.15$&$\gamma\gamma\to\pi^0\pi^0$
& \cite{oller08a} (result 5)\\
$3.9\pm 0.6$&$\gamma\gamma\to\pi^0\pi^0,\pi^+\pi^-$
& \cite{mennessier08} (result 6)\\
$3.1\pm 0.5$&$\gamma\gamma\to\pi^+\pi^-$
&\cite{pennington08} (result 7)\\
$2.4\pm 0.4$&$\gamma\gamma\to\pi^+\pi^-$
&\cite{pennington08} (result 8)\\
\hline
$2.3\pm 0.4$
& average of the results $(1-8)$&\\
\hline
\hline
\end{tabular}
\label{oller5}
\end{center}
\end{table}
The weighted average of the results 1--8 is
$\Gamma(\sigma\to\gamma\gamma)=(2.3\pm 0.4)$ keV and thus consistent with the
present result and also consistent  with the result of \cite{beveren09}.
We consider this general agreement as quite satisfactory.

\subsection{Properties of the reaction $\gamma\gamma\to \sigma\to N\bar{N}$ and
the $t$-channel part of the polarizability difference
$(\alpha-\beta)^t$}

\begin{figure}[h]
\centering\includegraphics[width=0.5\linewidth]{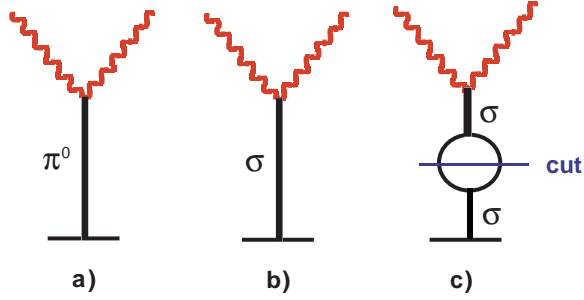}
\caption{$\pi^0$ and $\sigma$ pole graphs.}
\label{pi-sigma-pole}
\end{figure}

The pole at  $\sqrt{s_\sigma}=666$ MeV
has gained importance because it provides an easy access to the calculation
of  the $t$-channel part of Compton scattering as given by the   process
 $\gamma\gamma\to\sigma\to N\bar{N}$. This is illustrated in
Figure \ref{pi-sigma-pole}. In a) and b) the 
$t$-channel pole contributions corresponding to the $\pi^0$ and $\sigma$
meson  are compared with each other. These two
poles differ by  their locations at $m_{\pi^0}=135.0$ MeV 
and $m_\sigma= 666$ MeV, respectively, and their two-photon widths. 
In addition they differ by the fact that the 
pseudoscalar meson couples to two photons with perpendicular planes of
linear polarization whereas the scalar meson couples to two photons
with parallel planes of linear polarization. Except for this there are no
further differences in the two $t$-channel contributions. 
The $\sigma$ meson pole contribution was originally used in \cite{lvov97}
in connection with the prediction of differential cross sections
for Compton scattering. Its relation to the QLL$\sigma$M  was first
described in \cite{schumacher06}. Prior to this discovery \cite{schumacher06}
the calculation of the scalar $t$-channel contribution to nucleon
Compton scattering had to rely on the available information on the
two reactions $\gamma\to\sigma\to\pi\pi$ and   $\pi\pi\to\sigma\to N\bar{N}$
where the $\sigma$ meson was taken into account through the $\pi\pi$
phase relation. This corresponds to the graph c) in Figure \ref{pi-sigma-pole}.
Here the $\sigma$ propagator  has to be cut into two pieces for the 
purpose of the
calculation. This procedure has also been applied by BEFT 
\cite{bernabeu74}
to make predictions for the $t$-channel contribution of the difference    
$(\alpha-\beta)^t$ of the electric and magnetic polarizability.

In more detail the procedures of calculating $(\alpha-\beta)^t$ from the
graphs b) and c) in Figure \ref{pi-sigma-pole} are given in Eqs. 
 (\ref{BackSR1}) and (\ref{BackSR2}), respectively. Using the 
$\gamma\gamma\to\sigma \to N{\bar N}$ reaction the result 
\begin{equation}
(\alpha-\beta)^t=\frac{{\cal M}(\sigma\to\gamma\gamma)\,
g_{\sigma NN}}{2\pi \, m^2_\sigma }=15.2
\label{BackSR1}
\end{equation}
(in units of 10$^{-4}$fm$^3$) is obtained when inserting 
${\cal M}(\sigma\to\gamma\gamma)=(5\alpha_e)/(3\pi f_\pi)$, $m_\sigma=666$ MeV
and $g_{\sigma NN}=13.169\pm 0.057$ \cite{bugg04}.
In case of graph  c) in Figure \ref{pi-sigma-pole} we restrict ourselves 
in the calculation
of the $t$-channel absorptive part to intermediate states with two
pions with angular momentum $J\leq 2$. Then the $t$-channel part of 
the BEFT sum rule \cite{bernabeu74}
takes the form:
\begin{eqnarray}
(\alpha-\beta)^t&=& 
 \frac{1}{16 \pi^2}\int^\infty_{4\, m^2_\pi}\frac{dt}{t^2}\frac{16}{4\,m^2-t}
\left(\frac{t-4\,m^2_\pi}{t}\right)^{1/2}\Big[f^0_+(t)
  F^{0*}_{0}(t)\nonumber\\
&&-\left(m^2-\frac{t}{4}\right)\left(\frac{t}{4}-m^2_\pi\right)
f^2_+(t) F^{2*}_{0}(t)\Big],\label{BackSR2}
\end{eqnarray}
where
$f^{(0,2)}_+(t)$ and $F^{(0,2)}_0(t)$ are the partial-wave helicity
amplitudes of the processes $N\bar{N}\to \pi\pi$ and 
$\pi\pi\to \gamma\gamma$ with angular momentum $J=0$ and $2$,
respectively, and isospin $I=0$.  Though the quantities entering into
Eq. (\ref{BackSR2}) have a clear-cut definition the calculation is not easy.
This is reflected by the long history of different 
approaches to get reliable numbers
for $(\alpha-\beta)^t$ based on this equation. A description of these
approaches is given  in \cite{schumacher05}.

Very recently \cite{bernabeu08}
the experimental value of $(\alpha-\beta)^t$ has been used to make a
prediction for  the two-photon width 
$\Gamma(\sigma\to\gamma\gamma)$ of the $\sigma$ meson using Eq. 
(\ref{BackSR2}), with the $\sigma$ meson being a pole  at 441-i\,272 MeV
on the second second Riemann sheet. The result of this
latter calculation is 
$\Gamma_{\rm pole}(\sigma\to\gamma\gamma)=(1.2\pm 0.4)$ keV 
which still is compatible  with  the range of data listed in 
Table \ref{oller5} and, therefore, is  in line with the suggestion made 
by Figure \ref{pi-sigma-pole} that the pole representation leads to a good
approximation  of the scalar $t$-channel contribution.

\section{Compton scattering and polarizabilities 
of the nucleon \label{compton}}

The differential cross section for Compton scattering
may be written in the form \cite{babusci98} 
\begin{equation}
\frac{d\sigma}{d\Omega}=\Phi^2|T_{fi}|^2
\label{compton2}
\end{equation}
with $\Phi=\frac{1}{8\pi m}\frac{\omega'}{\omega}$
in the laboratory frame and $\Phi=\frac{1}{8\pi \sqrt{s}}$ in the c.m.
frame. The quantity $m$ is the mass of the nucleon, $\omega$ the energy of the
incoming photon,  $\omega'$ the  energy 
of the outgoing photon in the laboratory frame
and $\sqrt{s}$
the total energy. For the following discussion it is convenient to use the 
laboratory frame and to consider special cases for the amplitude
$T_{fi}$. These
special cases are the extreme forward $(\theta=0)$ and extreme backward
$(\theta=\pi)$ direction where the amplitudes for Compton scattering may be
written in the form \cite{babusci98}
\begin{eqnarray}
&&\frac{1}{8\pi m}[T_{fi}]_{\theta=0}=f_0(\omega)\veceps'^*\cdot\veceps
+g_0(\omega)i\,\sigbf\cdot(\veceps'^*\times\veceps),\label{compton3}\\
&&\frac{1}{8\pi m}[T_{fi}]_{\theta=\pi}=f_\pi(\omega)\veceps'^*\cdot\veceps
+g_\pi(\omega)i\,\sigbf\cdot(\veceps'^*\times\veceps)\label{compton4}.
\end{eqnarray}
In (\ref{compton3}) $f_0(\omega)$ is the forward scattering amplitude
with the two photons in parallel planes of linear polarization, 
$g_0(\omega)$ the forward scattering amplitude   with the two photons
in perpendicular planes of linear polarization, 
$f_\pi(\omega)$ the backward scattering amplitude 
with the two photons in parallel planes of linear polarization and 
$g_\pi(\omega)$ the backward scattering amplitude 
with the two photons in perpendicular planes of linear polarization. 

Following Babusci et al. \cite{babusci98} the equations (\ref{compton3})
and (\ref{compton4}) can be used to define the electromagnetic
polarizabilities and spin-polarizabilities as the lowest-order
coefficients in an $\omega$-dependent development of the
nucleon-structure dependent parts of the scattering amplitudes:
 \begin{eqnarray}
f_0(\omega) & = & - ({e^2}/{4 \pi\, m})\,q^2 + 
{\omega}^2 ({\alpha}
+{\beta}) + {\cal O}({\omega}^4), \label{f0}\\
g_0( \omega) &=&  \omega\left[ - ({e^2}/{8 \pi\, m^2})\,   
{\kappa}^2 
+ {\omega}^2
{\gamma_0}  + {\cal O}({\omega}^4) \right], \label{g0}\\
f_\pi(\omega) &=& \left(1+({\omega'\omega}/{m^2})\right)^{1/2} 
[-({e^2}/{4 \pi\, m})\,q^2 +       
\omega\omega'({\alpha} - {\beta}) 
+{\cal O}({\omega}^2{{\omega}'}^2)], \label{fpi}\\
g_\pi(\omega) &=& \sqrt{\omega\omega'}[
({e^2}/{8 \pi\, m^2})  
( {\kappa}^2 + 4q
{\kappa} + 2q^2)
+ \omega\omega'
{{\gamma_\pi}} + {\cal O}
({\omega}^2 {{\omega}'}^2)],  \label{gpi}
\end{eqnarray}
where $q e$ is the electric charge ($e^2/4\pi=1/137.04$), 
$\kappa$ the anomalous magnetic
moment of the nucleon and $\omega'=\omega/(1+\frac{2\omega}{m})$.

In the relations for $f_0(\omega)$ and $f_\pi(\omega)$ the first
nucleon structure dependent coefficients are the photon-helicity non-flip 
$(\alpha+\beta)$ and photon-helicity flip $(\alpha-\beta)$ linear
combinations of the electromagnetic polarizabilities $\alpha$ and
$\beta$. In the relations for $g_0(\omega)$ and $g_\pi(\omega)$
the corresponding coefficients are the spin polarizabilities
$\gamma_0$ and $\gamma_\pi$, respectively. 
The relations between 
the amplitudes $f$ and $g$ 
and the invariant amplitudes $A_i$ \cite{babusci98,lvov97}
 are
\begin{eqnarray}
&&f_0(\omega)= -\frac{\omega^2}{2\pi}\left[A_3(\nu,t)+ A_6(\nu,t)
\right],\quad\quad\quad
g_0(\omega)=\frac{\omega^3}{2\pi\, m}A_4(\nu,t), \label{T3}\\
&&f_\pi(\omega)=-\frac{\omega\omega'}{2\pi}\left(1+\frac{\omega\omega'}
{m^2}\right)^{1/2}\left[ 
A_1(\nu,t) - \frac{t}{4 m^2}A_5(\nu,t)\right],\label{T4}\\
&&g_\pi(\omega)=-\frac{\omega\omega'}{2\pi m}\sqrt{\omega\omega'}
\left[
A_2(\nu,t)+ \left(1-\frac{t}{4 m^2}\right)A_5(\nu,t)\right]
,\label{T5}\\
&&\omega'(\theta=\pi)=\frac{\omega}{1+2\frac{\omega}{m}},\,\,
\nu=\frac12 (\omega+\omega'),\,\, t(\theta=0)=0,
\,\,t(\theta=\pi)=-4\omega\omega.'
\label{T6}
\end{eqnarray}

For the electric, $\alpha$, and magnetic, $\beta$,  polarizabilities 
and the spin polarizabilities $\gamma_0$ and $\gamma_\pi$ for the
forward and backward directions, respectively, 
we obtain the relations
\begin{eqnarray}
&&\alpha+\beta = -\frac{1}{2\pi}\left[A^{\rm nB}_3(0,0)+ 
A^{\rm nB}_6(0,0)\right], \quad 
\alpha-\beta = -\frac{1}{2\pi}
\left[A^{\rm nB}_1(0,0)\right], \nonumber \\ 
&&\gamma_0= \frac{1}{2\pi m}\left[A^{\rm nB}_4(0,0)
\right], \quad\quad\quad\quad \quad\quad\quad\,\,\,
\gamma_\pi = -\frac{1}{2\pi m}
\left[A^{\rm nB}_2(0,0)+A^{\rm nB}_5(0,0) \right],
\label{T7}
\end{eqnarray}
where $A_i^{\rm nB}$ are the non-Born parts of the invariant amplitudes.

\subsection{Compton scattering and electromagnetic fields}

Polarizabilities may be measured by simultaneous interaction
of two photons with the nucleon. In case of static fields this may be written
in the form
\begin{equation}
H^{(2)}= -\frac12 \,4\,\pi\,\alpha \,{\bf E}^2
-\frac12 \,4\,\pi\,\beta \,{\bf H}^2,
\label{staticfields}
\end{equation}
where  the quantity $H^{(2)}$ is the energy change in the
electric and magnetic fields due to the polarizabilities. The first part 
of the r.h.s. of Eq.  
(\ref{staticfields}) is realized in experiments where slow neutrons
are scattered in the electric field of a heavy nucleus, leading to a
measurement of the electric polarizability.  
\begin{figure}[h]
\centering\includegraphics[width=0.4\linewidth]{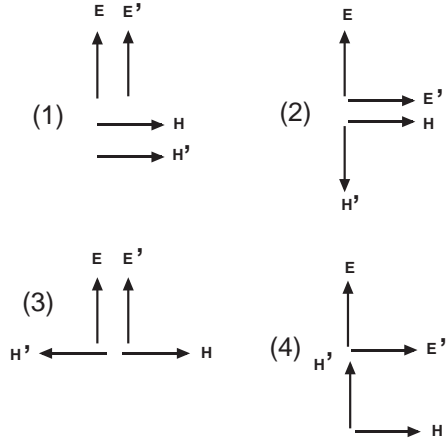}
\caption{Compton scattering viewed as simultaneous interaction of two electric
field vectors ${\bf E},{\bf E'}$ and  magnetic field vectors 
${\bf H},{\bf H'}$ for four different
cases. (1) Helicity independent forward Compton scattering as given by the 
amplitude $f_0(\omega)$.  
(2) Helicity dependent forward Compton scattering as given by the 
amplitude $g_0(\omega)$.
(3) Helicity independent backward  Compton scattering as given by the 
amplitude $f_\pi(\omega)$.
(4) Helicity dependent backward  Compton scattering as given by the 
amplitude $g_\pi(\omega)$. Longitudinal photons can only provide two
electric vectors with parallel planes of linear polarization as shown
in the upper part of panel (1). In panels (2) and (4) the direction of
rotation leading from ${\bf E}$ to  ${\bf E}'$ depends on the helicity
difference $|\lambda_p-\lambda_\gamma|$.
}
\label{EHfields}
\end{figure}
Compton scattering in the forward and backward  directions lead to  more 
general combinations of electric and magnetic fields. This is depicted in
Figure  \ref{EHfields}. Panel (1) corresponds to the amplitude $f_0(\omega)$,
i.e. to Compton scattering in the forward direction with parallel electric
and magnetic fields. This case can be compared with Eq. (\ref{staticfields}).
Electromagnetic scattering of neutrons corresponds  to the
upper part of panel (1) where two electric vectors are shown. The cases
described in panels (2) -- (4) can be realized with real photons only.
Panel (2) corresponds to helicity dependent Compton scattering in the forward
direction as described by the amplitude $g_0(\omega)$. Panel (3) corresponds
to the amplitude $f_\pi(\omega)$ and panel (4) 
to the amplitude $g_\pi(\omega)$. 
According to the definitions given in Eq. (\ref{T7}),
 panel (3) corresponds to the case where a $\sigma$ meson while attached
to a constituent quark interacts with the two photons and panel (4)
 to the case where a $\pi^0$ meson while attached
to a constituent quark interacts with the two photons.

We have argued that electromagnetic scattering of neutrons corresponds
to the upper part of panel (1). Furthermore, we have argued that 
the reaction $\gamma\gamma\to\sigma\to N\bar{N}$ makes a contribution
to the case of panel (3). At a first sight there may be a contradiction
to the observation that electromagnetic scattering of neutrons and 
Compton scattering measure the same electric polarizability. The explanation
that there is no contradiction
is as follows. In terms of electric and magnetic fields the $\sigma$ meson
always
makes a contribution if the two field  vectors are either parallel or
antiparallel. However, with $\alpha(\sigma) = - \beta(\sigma)$ these two
parts cancel in the case of panel (1) for real photons but do not cancel
for virtual photons at low neutron velocities where magnetic fields 
are absent.

\subsection{Composition of the Compton scattering amplitudes}

\begin{figure}[h]
\begin{center}
\includegraphics[width=0.5\linewidth]{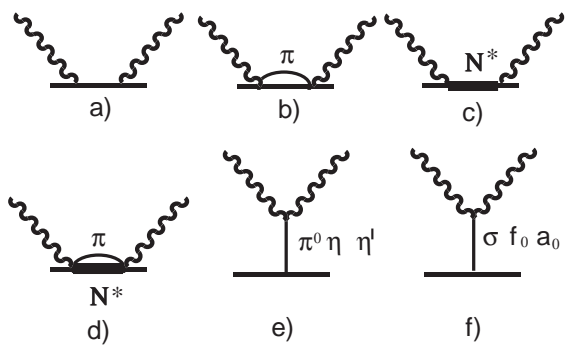}
\end{center}
\caption{Graphs of processes contributing  to Compton scattering by
  the nucleon$^{3}$. a) Born terms, b) Compton scattering via nonresonant
$1\pi$ excitation of the nucleon, c) Compton scattering  via excitation of the
resonant states of the nucleon, d) summary of other nonresonant or partly
resonant excitation processes of the nucleon contributing to Compton
scattering. The corresponding crossed graphs have been omitted. e)
pseudoscalar $t$-channel exchange, f) scalar $t$-channel exchange.  Regge
pole exchanges contributing in the forward direction are not explicitly shown.}
\label{comp-graph}
\end{figure}

The total Compton differential cross section is represented via the
graphs\footnote{Please note that the graphs are shown for
    illustration only whereas the calculations are based on dispersion theory.
\label{feyn}}
of Figure \ref{comp-graph}.
Compton scattering takes place via Born terms a) without internal excitation 
of the nucleon and non-Born terms  b) to f) which make contributions to the
polarizabilities. The graphs corresponding to the $s$-channel, {\it viz.}
a) to d), have to be supplemented by crossed graphs which are not shown here.
Graph f) is of special interest because it represents the scalar $t$-channel
contribution. In addition to the contribution of the
$\sigma$ meson there are small contributions
of the $f_0(980)$ and $a_0(980)$ mesons. 
\begin{figure}[ht]
\centering\includegraphics[width=0.4\linewidth]{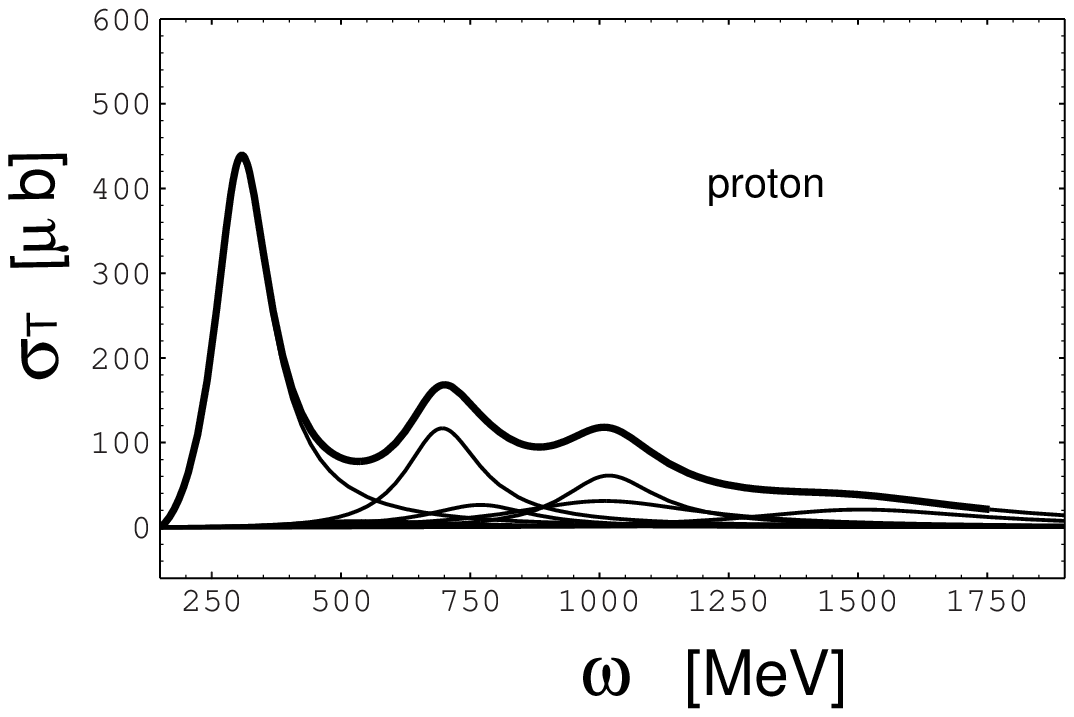}
\centering\includegraphics[width=0.4\linewidth]{{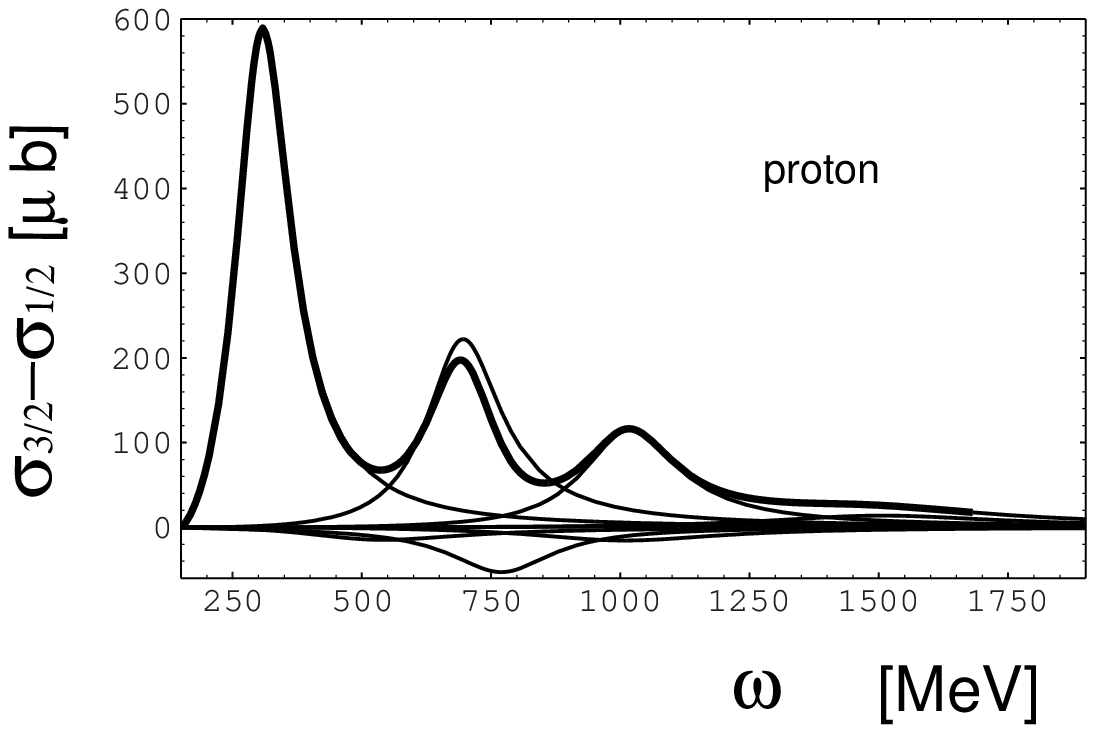}}
\caption{Resonant part of the total photo absorption cross section for 
$\sigma_T=\frac12 (\sigma_{1/2}+\sigma_{3/2})$ and $\sigma_{3/2}-\sigma_{1/2}$.
Thick line: Sum of all resonances. Thin lines:
Contributing single resonances.
}
\label{coordinates}
\end{figure} 
Figure \ref{coordinates} gives an overview over the spin-independent
(left panel) and spin-dependent (right panel) contributions to graph c) 
of Figure \ref{comp-graph}. The cross sections are dominated by the 
$P_{33}(1232)$, $D_{13}(1520)$ and $F_{15}(1680)$ resonances. The main
difference between the spin-independent cross section (left panel) and the
spin-dependent cross section (right panel) is the enhancement 
of the $P_{33}(1232)$ contribution by a factor 1.26 which can be traced back
to the E2/M1 ratio of the  $P_{33}(1232)$ resonance \cite{schumacher09}.
Without the multipolarity mixing the two resonance curves would be the same
for the $P_{33}(1232)$ resonance.
\begin{table}[h]
\caption{Predicted polarizabilities for the proton compared with 
experimental data. The unit is 10$^{-4}$fm$^3$ (for details
see \cite{schumacher09})}
\begin{center}
\begin{tabular}{lll}
\hline
&$\alpha_p$ & $\beta_p$ \\
\hline
$P_{33}(1232)$ & $-1.1$ & +8.3 \\
other resonances &+1.1& +0.2\\
$E_{0+}$ nonresonant &+3.2 & $-0.3$\\
other nonresonant &+1.3 & +1.2\\
$\sigma$ $t$-channel& +7.6 & $-7.6$\\
$f_0$, $a_0$  $t$-channel& $-0.1$ & +0.1\\
sum & +12.0& +1.9\\
\hline
experiment& $+12.0\pm 0.6$ & $+1.9\mp 0.6$\\
\hline\hline
\end{tabular}
\end{center}
\label{tab}
\end{table}
The nonresonant cross section  corresponding to graph b) in Figure 
\ref{comp-graph} has contributions from the single-pion $E_{0+}$,
$(M,E)^{(1/2)}_{1+}$ and $M^{(3/2)}_{1-}$ CGLN amplitudes with the 
 $E_{0+}$ contribution being the largest.
The polarizabilities of the nucleon have been measured with high precision
and compared with predictions from dispersion theory in several
investigations. The most recent one is published in \cite{schumacher09}. 
Therefore, it is not necessary to give an extensive coverage of the
polarizabilities here but it is sufficient to present  the essential
results.
The $s$-channel contributions $\alpha^s$ and $\beta^s$ of the electromagnetic
polarizabilities have been calculated from the multipole content of the 
photoabsorption cross section whereas the  $t$-channel contribution
is given by the transition matrix element 
${\cal M}(\sigma\to\gamma\gamma)$ of Eq. (\ref{twoph}), 
the $\sigma$-nucleon coupling constant
$g_{\pi NN}=g_{\sigma NN}= 13.169\pm 0.057$ \cite{bugg04} and 
the $\sigma$-meson mass $m_\sigma= 666$
MeV via $\alpha^t=-\beta^t={\cal M}(\sigma\to\gamma\gamma)\,g_{\sigma
  NN}/4\pi m^2_\sigma=7.6$. Predictions for partial contributions to the
polarizabilities of the proton are given in Table \ref{tab}.
The purpose of Table \ref{tab} is to show that the experimental data are
well understood in terms of excitation processes of the nucleon and that
the $\sigma$-meson $t$-channel makes a dominant contribution.

From Table \ref{tab} the following conclusions may be drawn:
(i) the prediction obtained for the
two-photon width $\Gamma(\sigma\to\gamma\gamma)$ from the QLL$\sigma$M
has been confirmed with a high level of precision through the excellent
agreement of the experimental electric polarizability $\alpha_p$ of the proton
with the corresponding predicted quantity. (ii) an error for the
prediction of $\Gamma(\sigma\to\gamma\gamma)$ may be obtained from the
experimental error of $\alpha_p$ being $\Delta\alpha_p/\alpha_p\approx
5\% $. Since the $\sigma$ meson contribution to the electric polarizability 
is by far the largest it is possible to relate the errors of $\alpha_p$
and $\Gamma(\sigma\to\gamma\gamma)$ to each other.
Using 
$\Delta\Gamma(\sigma\to\gamma\gamma)/\Gamma(\sigma\to\gamma\gamma)
\approx  2 \Delta\alpha_p/\alpha_p\approx 10\%$ the error given in Eq.
(\ref{width}) is obtained.

\section{Observation of
the $\sigma$-meson contribution
to the differential cross section for Compton scattering} 

The BEFT sum rule \cite{bernabeu74} may be derived from the non-Born (nB)
part of the invariant amplitude
\begin{equation}
{\tilde A}_1(s,u,t)\equiv A_1(s,u,t)-\frac{t}{4\,m^2}A_5(s,u,t)
\label{tildeA}
\end{equation}
by applying the fixed-$\theta$ dispersion relation for $\theta=\pi$
\begin{eqnarray}
&&{\rm Re}{\tilde A}^{\rm nB}_1=\frac{1}{\pi}{\cal P}\int^\infty_{s_0}
\left(\frac{1}{s'-s}+\frac{1}{s'-u}-\frac{1}{s'}\right){\rm Im}_s{\tilde A}_1
(s',u',t')\,ds'\nonumber\\
&&\quad\quad\quad\quad +\frac{1}{\pi}{\cal P}\int^\infty_{t_0}\,
{\rm Im}_t{\tilde A}_1(s',u',t')\frac{dt'}{t'-t} \label{dispersion}
\end{eqnarray} 
with $su=m^4$. Then the difference of the electromagnetic polarizabilities is
given by 
\begin{equation}
\alpha-\beta=-\frac{1}{2\pi}{\rm Re}{\tilde A}^{\rm nB}_1(s=m^2,u=m^2,t=0)=
(\alpha-\beta)^s+(\alpha-\beta)^t. \label{defalphaminusbeta}
\end{equation}
A calculation analogous to the derivation of the BEFT sum rule
(see e.g. \cite{schumacher05})
but for general $s$ and $t$ and for the pole representation of the
$t$-channel  part leads to a  generalized BEFT sum rule in the form
\begin{eqnarray}
&&(\alpha-\beta)^s(\omega)=\frac{1}{2\pi^2}{\cal P}\int^\infty_{\omega_0}
\sqrt{1+2\frac{\omega'}{m}}\,\,\,\frac{\omega'^3+\omega'^2 (m/2)}{\omega'^3
+\omega'^2 a(\omega) +\omega' b(\omega)+ c(\omega)}\,\,\sigma_{f_\pi}(\omega')
\,\,\frac{d\omega'}{\omega'^2} \label{BEFT1}\\
&&(\alpha-\beta)^t(\omega)=-\frac{1}{2\pi}\,\,\frac{{\cal M}
(\sigma\to\gamma\gamma)\,
g_{\sigma NN} }{t(\omega)-m^2_\sigma}  \label{BEFT2}
\end{eqnarray}
with $\sigma_{f_\pi}(\omega)=\sigma(\omega,E1,\,\,M2,\cdots)-
\sigma(\omega,M1,\,\,E2,\cdots)$ and
\begin{equation}
a(\omega)=\frac{-4\omega^2 +2m\omega+m^2}{2(m+2\omega)},\,\,\,
b(\omega)=-\frac{2m\omega^2}{m+2\omega},\,\,\, c(\omega)=-\frac{m^2\omega^2}{
2(m+2\omega)},\,\,\, t(\omega)=-4\frac{\omega^2}{1+2\frac{\omega}{m}}.
\label{BEFT3}
\end{equation}
It should be noted that Eq. (\ref{BEFT2}) follows from the reaction
$\gamma\gamma\to \sigma \to N\bar{N}$ where $\sigma$ is a genuine $q\bar{q}$ 
state. This means that the $\sigma$ meson 
has a definite mass of $m_\sigma= 666$ MeV
as predicted by the QLL$\sigma$M. A complex mass parameter or a mass
distribution are excluded as explained in the text following Eq. (\ref{9}).

\begin{figure}[h]
\centering\includegraphics[width=0.4\linewidth]{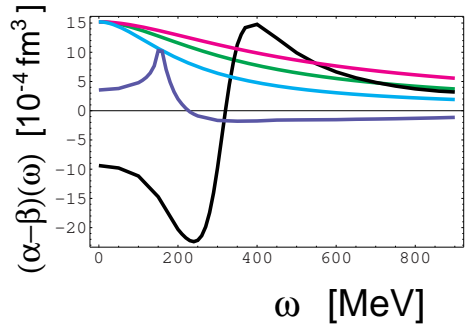}
\caption{Difference of generalized polarizabilities $(\alpha-\beta)(\omega)$
versus photon energy $\omega$. Solid curve starting at
$(\alpha-\beta)(0)=-9.4$: contribution of the $P_{33}(1232)$ resonance.
Solid  curve starting at $(\alpha-\beta)(0)=+ 3.5$: Contribution of the
nonresonant $E_{0+}$ amplitude. Curves starting at 
$(\alpha-\beta)(0)=+ 15.2$: $t$-channel contribution of the $\sigma$-meson
calculated for different $\sigma$-meson masses. Upper curve (dark grey or red):
$m_\sigma= 800$ MeV. Middle curve (grey or green):  $m_\sigma= 600$ MeV.
Lower curve (light grey or blue):  $m_\sigma= 400$ MeV.
Not shown is the contribution of the $D_{13}(1520)$ resonance which cancels
the $E_{0+}$ contribution in the relevant energy range from 400 to 700 MeV.
}
\label{aminusb}
\end{figure}

The generalized polarizabilities as defined in (\ref{BEFT1}) and
(\ref{BEFT2}) are depicted in Figure $\ref{aminusb}$. The solid curve starting
at $(\alpha-\beta)(0)=-9.4$ represents the contribution of the $P_{33}(1232)$
resonance.
The solid  curve starting at 
$(\alpha-\beta)(0)=+ 3.5$ represents the contribution of the
nonresonant $E_{0+}$ amplitude. The curves  starting at 
$(\alpha-\beta)(0)=+ 15.2$ represent the  $t$-channel contribution of the 
$\sigma$-meson calculated for different $\sigma$-meson masses. 
The upper curve (dark grey or red) has been calculated for
$m_\sigma= 800$ MeV, the middle curve (grey or green) for   $m_\sigma= 600$ MeV
and the lower curve (light grey of blue) for  $m_\sigma= 400$ MeV. In principle
 also a contribution of the $D_{13}(1520)$ resonance has to be taken unto
account. This contribution is not shown in order not to overload the figure.
The effect of this contribution is to cancel the $E_{0+}$ contribution in the
energy range from 400 to 700 MeV  which is the most relevant part of the
spectrum for the comparison with 
\begin{figure}[h]
\centering\includegraphics[width=0.6\linewidth]{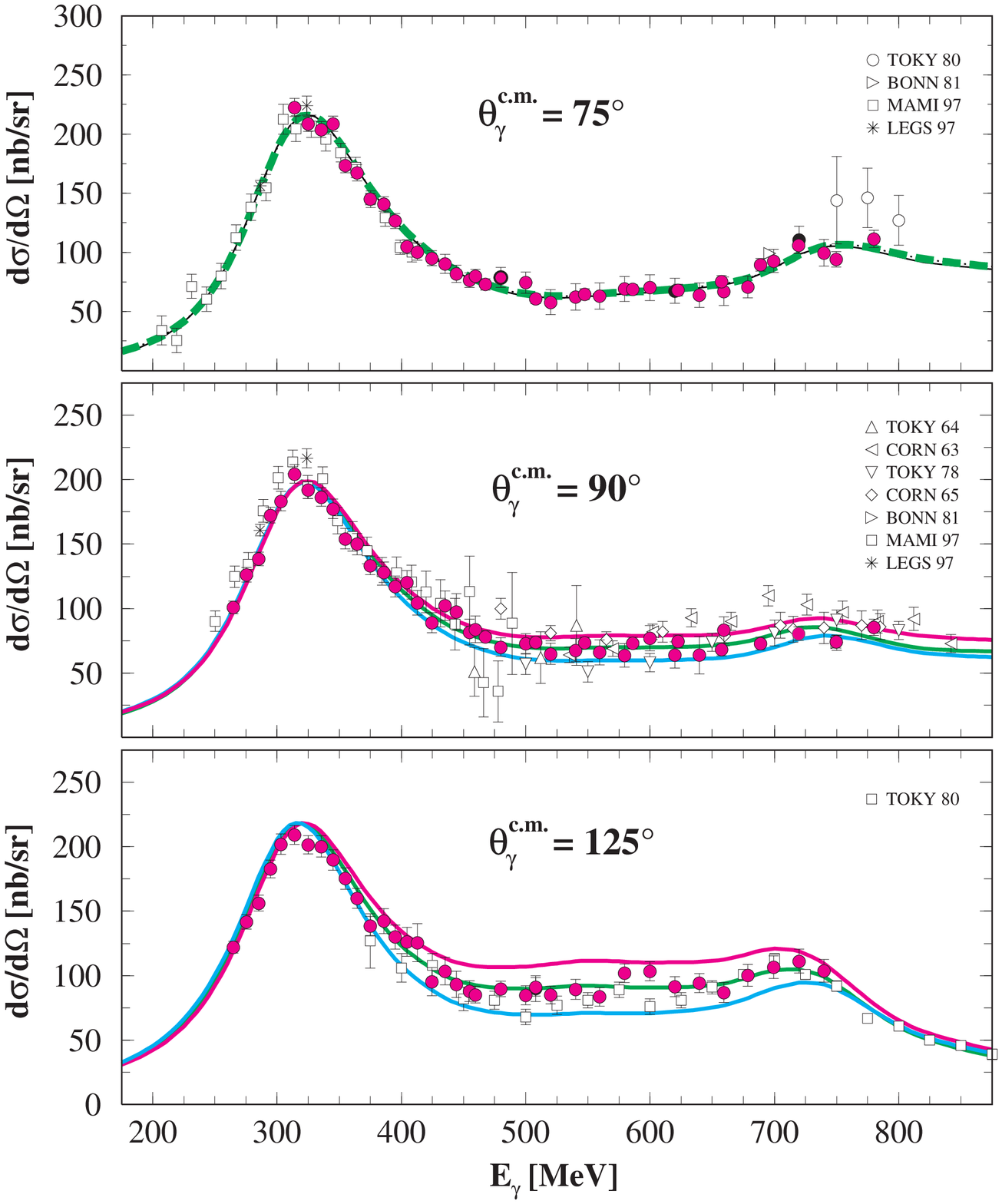}
\caption{Differential cross sections for Compton scattering by the proton
  versus photon energy. The three panels contain data corresponding to the
  c.m.-angles of
$75^0$, $90^0$ and $125^0$. The three curves are calculated for different mass
  parameters $m_\sigma=800$ MeV (upper, dark grey or red), 600 MeV 
(center, grey or green) and 400 MeV (lower, light grey or blue).
}
\label{fig9}
\end{figure}
experimental
data as discussed later. It is apparent that
the contribution of the $\sigma$ meson enters via a constructive interference
with the contribution of the $P_{33}(1232)$ resonance. This strongly enhances
the effects of the $\sigma$ meson in the total differential cross section
for Compton scattering at large scattering angles in the range between 400 and
700 MeV.

Differential cross sections for Compton scattering by the proton have been
measured at the high duty-factor electron facility MAMI (Mainz)
 \cite{galler01,wolf01}. In this
experiment it was possible to cover  the total energy range from
250 MeV to 800 MeV and angular range in the laboratory from $30^\circ$
to $150^\circ$ in one experimental run. The result of this experiment is shown
in Figure \ref{fig9}.
The experimental data  collected in angular intervals are
given for the c.m. scattering angles $75^\circ$,
$90^\circ$ and $125^\circ$ and compared 
with predictions based on dispersion
theory. As in Figure  \ref{aminusb} the mass of the $\sigma$ meson has been
varied using  800 MeV, 600 MeV and 400 MeV while keeping the quantity
$(\alpha-\beta)(0)$ constant. From kinematical reasons effects of the 
$\sigma$ meson in the
differential cross section for Compton are only expected in the backward
direction and are largest at $\theta=180^\circ$. This is in agreement with the
observations made in Figure \ref{fig9}. At $\theta^{\rm c.m.}=125^\circ$
there are three different curves visible in the energy range between 400 MeV
and 700 MeV corresponding to the different $\sigma$ masses: 800 MeV (upper),
600 MeV (center) and 400 MeV (lower). This apparently 
is in a complete agreement with our
expectation from Figure \ref{aminusb}. This means that we have seen the
contribution from the $\sigma$ meson pole in the experimental
differential cross section. Furthermore,
the mass we have determined is
$m_\sigma\approx 600$ MeV in agreement with  the prediction made by the
QLL$\sigma$M for the  reaction
 $\gamma\gamma\to \sigma \to N \bar{N}$, {\it viz.} $m_\sigma=666$ MeV.
This mass corresponds to the location of the pole in the $t$-channel
of Compton scattering as outlined above.

\section{Discussion \label{discussion}}

In the present  paper we have shown that the evaluation of the reaction
$\gamma\gamma\to  \sigma\to N{\bar N}$ with properties of the
$\sigma$ meson as given by  the QLL$\sigma$M
leads to   predictions for the electromagnetic 
polarizabilities of the nucleon being valid
  with a high
level of precision. This strongly confirms  the supposition that the $\sigma$
meson indeed enters into the polarizabilities via the 
$\gamma\gamma\to  \sigma\to N{\bar N}$ process with the $\sigma$ meson having
a mass of $m_\sigma= 666$ MeV.  An even more direct observation of the
$\gamma\gamma\to  \sigma\to N{\bar N}$ reaction
has also been made through its  large
contribution  to the differential cross section for Compton scattering by the
proton in the energy range from $\omega=400$ MeV to 700 MeV 
where such a large contribution is expected. 
 On the quark level this
process may be understood as Compton  scattering by the $\sigma$ meson while
being a part of the constituent quark. 

Prior to the discovery \cite{lvov97,schumacher06}
of the $\sigma$ meson pole representation of the scalar-isoscalar
$t$-channel of Compton scattering
with the $\sigma$-meson having a mass of $m_\sigma=666$ MeV, it was only
possible to construct this quantity  from  a combination of the two
reactions $\gamma\gamma\to\pi\pi$ and $\pi\pi\to N{\bar N}$ with the $\sigma$
meson being described via a pole on the second Riemann sheet
at $\sqrt{s_\sigma}=(441- i\,272)$ MeV. Both procedures are correct and
equivalent to each other. But certainly the calculation on the basis
of Eq. (\ref{BackSR1}) is easier to perform than the calculation on the 
basis of Eq. (\ref{BackSR2}). 
A further result of the present investigation is that rather firm arguments
are obtained for the  two-photon width of the $\sigma$-meson being
$\Gamma(\sigma\to\gamma\gamma)=(2.6\pm 0.3)$ keV, thus removing a large
systematic uncertainty left over by different recent evaluations 
\cite{pennington06b,bernabeu08,oller08,oller08a,mennessier08,pennington08}
of the
$\sigma$-meson pole    at $\sqrt{s_\sigma}=(441- i\,272)$ MeV.

The $\sigma$ meson was originally introduced by Schwinger \cite{schwinger57}
in his general  attempt  to
explain symmetry breaking including also  the electro-weak (EW) 
sector. Later on it was shown
by Gell-Mann--Levy \cite{gellmann60}
that the $\sigma$ meson may be used to explain
the mass of the nucleon, whereas Higgs later on discussed the
EW  symmetry breaking in a form which nowadays is generally
accepted \cite{higgs64} and which enters into the standard model (SM).  
Similarities between
the $\sigma$ meson and the Higgs boson possibly may go beyond this 
interchange of names. 
Since the $\sigma$ meson is well understood on a quark level there have
been attempts to use it as a guide for models of the Higgs boson (see e.g. 
\cite{scadron06,fang97} and references therein).
 At present 
\cite{peters09} the SM Higgs boson is excluded at 95\% C.L. for a mass
lower than 114.4 GeV and for a mass in the range $160 < m_H < 170$ GeV.
This result is only valid for the SM Higgs boson and does not contain 
information about a non-SM Higgs boson  \cite{peters09}.

It is interesting to note that in a new development of ChPT the possible
role of the $f_0(600)$ meson  has been discussed  \cite{lensky09}.

\section*{Acknowledgment}

The author is indebted to J.A. Oller for providing valuable information
concerning his work.

\end{document}